\documentclass[]{spie}  

\usepackage{blindtext}
\usepackage{amsmath,amsfonts,amssymb}
\usepackage{graphicx}
\usepackage[colorlinks=true, allcolors=blue]{hyperref}
\usepackage{esvect}
\usepackage[utf8]{inputenc}
\usepackage{color}
\usepackage{wrapfig}
\usepackage{subfig}
\usepackage{indentfirst}
\usepackage{hhline}%
\usepackage{ltablex}
\usepackage{booktabs}
\usepackage{caption}
\usepackage{multirow}%
\usepackage{float}

\title{MRI Tumor Segmentation with Densely Connected 3D CNN}

\author[$\dagger$1]{Lele Chen}
\author[$\dagger$1]{Yue Wu}
\author[2]{Adora M. DSouza}
\author[3]{Anas Z. Abidin}
\author[2,3,4,5]{\\Axel~Wism\"{u}ller}
\author[1]{Chenliang~Xu}
\affil[1]{\normalsize Department of Computer Science, University of Rochester, NY, USA}
\affil[2]{Department of Electrical Engineering, University of Rochester, NY, USA}
\affil[3]{Department of Biomedical Engineering, University of Rochester, NY, USA}
\affil[4]{Department of Imaging Sciences, University of Rochester, NY, USA}
\affil[5]{Faculty of Medicine and Institute of Clinical Radiology, Ludwig Maximilian University, Germany}

\authorinfo{
$\dagger$These authors contributed equally.\\
Send correspondence to C. Xu~(chenliang.xu@rochester.edu).\\
Our code is available at \url{https://github.com/lelechen63/MRI-tumor-segmentation-Brats}.
}
\begin{document} 
\maketitle

\begin{abstract}
	
Glioma is one of the most common and aggressive types of primary brain tumors. The accurate segmentation of subcortical brain structures is crucial to the study of gliomas in that it helps the monitoring of the progression of gliomas and aids the evaluation of treatment outcomes. However, the large amount of required human labor makes it difficult to obtain the manually segmented Magnetic Resonance Imaging (MRI) data, limiting the use of precise quantitative measurements in the clinical practice. In this work, we try to address this problem by developing a 3D Convolutional Neural Network~(3D CNN) based model to automatically segment gliomas. The major difficulty of our segmentation model comes with the fact that the location, structure, and shape of gliomas vary significantly among different patients. In order to accurately classify each voxel, our model captures multi-scale contextual information by extracting features from two scales of receptive fields. To fully exploit the tumor structure, we propose a novel architecture that hierarchically segments different lesion regions of the necrotic and non-enhancing tumor~(NCR/NET), peritumoral edema~(ED) and GD-enhancing tumor~(ET). Additionally, we utilize densely connected convolutional blocks to further boost the performance. We train our model with a patch-wise training schema to mitigate the class imbalance problem. The proposed method is validated on the BraTS 2017 dataset~\cite{DBLP:journals/tmi/MenzeJBKFKBPSWL15} and it achieves Dice scores of 0.72, 0.83 and 0.81 for the complete tumor, tumor core and enhancing tumor, respectively. These results are comparable to the reported state-of-the-art results, and our method is better than existing 3D-based methods in terms of compactness, time and space efficiency. 

\end{abstract}


\section{INTRODUCTION}
\label{sec:intro}  
Glioma segmentation in MRI data provides valuable assistance for treatment planning, disease progression monitoring for oncological patients. Although deep learning technologies have shown a great potential in natural image segmentation, it is not straightforward to apply them for segmenting medical imaging data. In this work, we address the challenging problem of brain tumor segmentation using MRI scans. One of the major difficulties of this task is the class imbalance problem. Since the lesion areas in MRI scans can be extremely small in many cases, additional techniques need to be employed to avoid the background domination. A commonly used strategy is to extract small patches of the whole 3D volume with a pre-defined probability of being centered on lesion area, then train a Convolutional Neural Network (CNN) in a patch-wise fashion~\cite{DBLP:journals/corr/HavaeiDWBCBPJL15,DBLP:journals/mia/KamnitsasLNSKMR17,DBLP:journals/tmi/PereiraPAS16,DBLP:conf/miccai/ZhaoWSLFZ16,DBLP:journals/corr/BrebissonM15}. This approach is an effective way to keep the positive and negative samples balanced. 

Among the literature that utilizes this patch-wise training schema, two prediction strategies are widely used. The first one can be regarded as \textit{patch classification}~\cite{DBLP:journals/tmi/PereiraPAS16,DBLP:conf/miccai/ZhaoWSLFZ16,DBLP:journals/corr/BrebissonM15} where the model predicts a label for the central voxel of that patch. The second one is \textit{patch segmentation}~\cite{DBLP:journals/mia/KamnitsasLNSKMR17} 
where the model tries to get a dense prediction of the labels for multiple voxels in that patch simultaneously. In either case, the contextual information of voxels is of great importance to accurate predictions. Kamnitsas et al.~\cite{DBLP:journals/mia/KamnitsasLNSKMR17} make dense predictions on a shrunken feature map of each patch. Their CNN nodes with the receptive field lying outside the input patch are not predicted and do not contribute to the loss. The performance of their method is on par with those using patch classification schema, while it substantially improves the efficiency. In this work, we further improve the performance by introducing a multi-stage 3D CNN to make dense predictions based on two scales of receptive fields. Furthermore, our model is built upon densely connected blocks~\cite{huang2016densely} and utilizes a hierarchical architecture to consider different types of brain tumors. Next, we introduce the motivations for each technologies we have employed one-by-one. 

In CNN, the receptive field of a node is the region of the input signal that is involved in the multi-layer convolution up to that node in a forward pass. Thus, the size of the receptive field determines how much contextual information is taken into account. Some voxels may require only its neighboring information to generate the correct prediction, while others may need distant information. To make the model robust, we predict each voxel based on two sizes of receptive field in a multi-stage architecture. The classification scores are added to obtain the final score. We observe that the multi-stage design significantly boosts the performance. 

Residual learning~\cite{he2016deep}, since introduced, has gained popularity for all kinds of computer vision tasks. What sets it apart from traditional plain networks is the introduction of shortcut connections that enable the network to go much deeper~\cite{szegedy2015going}. 
The shortcut connections attenuate the notorious gradient vanishing problem since gradients can be back-propagated directly through shortcuts. From another point of view, shortcut connections make the entire network a huge ensemble of shallow networks, which boosts the performance. Recently, DenseNet~\cite{huang2016densely}, being a successor of applying shortcut connections, incorporates features from all previous layers (thus \textit{dense}). This change makes the network deeper while having a reduced number of parameters. It has led to improved effectiveness and efficiency in various image classification tasks. However, whether or not the dense connections can improve brain tumor segmentation remains unexplored. 

To fully exploit the hierarchical structure of the lesion area, we develop a novel architecture that hierarchically segments ED and its subregions, ET and NCR/NET. This is based on our observation that the ET and NCR/NET always lie inside of ED. We thus propose a two-pathway architecture to first segment ED with Fluid Attenuated Inversion Recovery (FLAIR) and T2-weighted (T2) scans, then segment ED and all subregions with all types of MRI scans.

We test our models on the Multimodal Brian Tumor Segmentation (BraTS)~\cite{DBLP:journals/tmi/MenzeJBKFKBPSWL15} 2017 benchmark. 
The experimental results on BraTS indicate that our model achieves comparable performance to the state-of-the-art methods without bells and whistles, which suggests the potential of our model being applied for lesion segmentation tasks in MRI scans. The innovations introduced by our paper can be summarized as follows: 1) we generalize densely connected blocks to 3D scenarios for MRI brain tumor segmentation tasks; 2) we introduce multi-scale receptive fields for accurate voxel classification and efficient dense inference; 3) we propose a hierarchical segmentation structure inspired by the structure of lesion regions.



\section{METHODS}
\label{sec:method}

\subsection{Hierarchical Segmentation Structure}
\label{sec:baseline}

\begin{figure}
\centering
\includegraphics[scale=0.4]{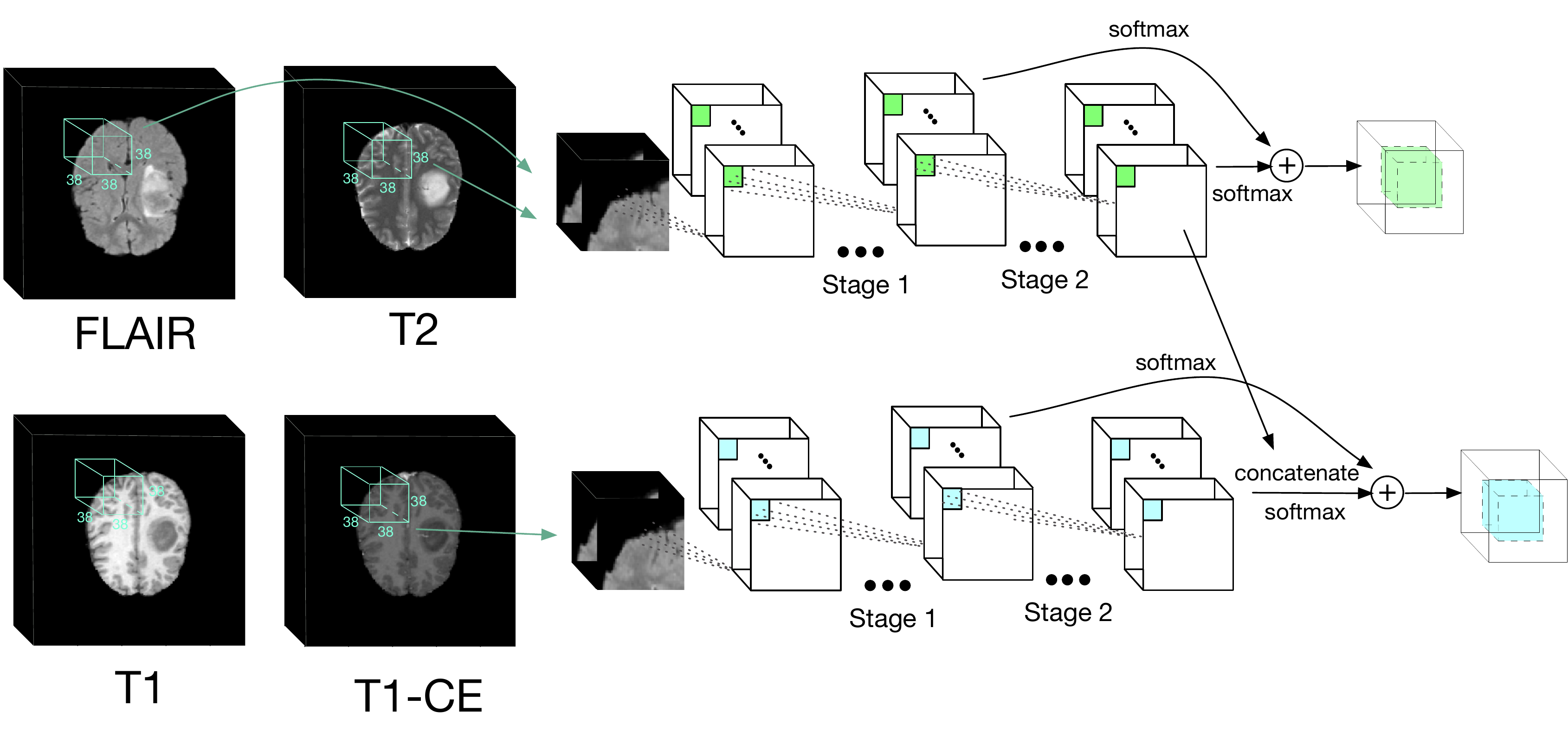}
\caption{The overview of our segmentation approach with densely connected 3D CNN hierarchical structure.}
\label{fig:overview} 
\end{figure} 

\begin{figure} [ht]
\centering
\includegraphics[scale=0.4]{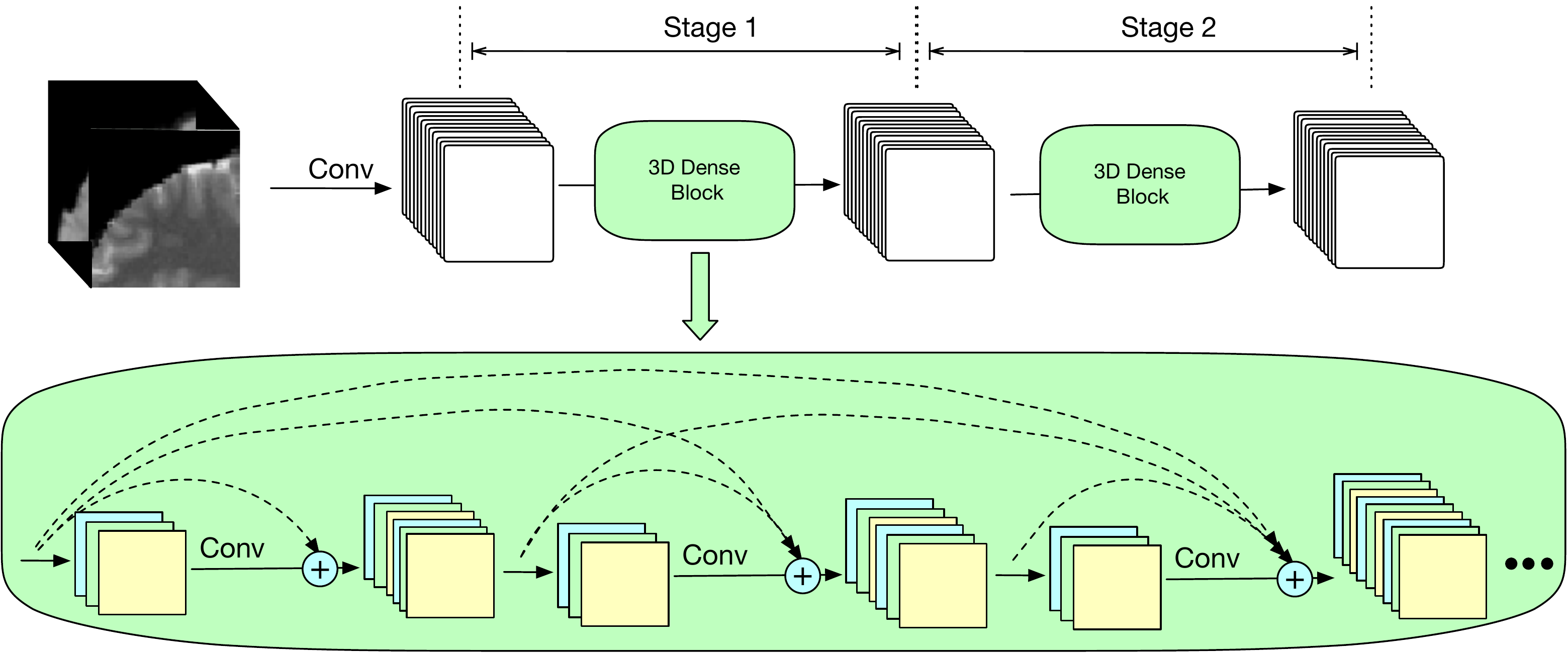}

\caption{The structure of densely connected 3D CNN feature extractor.}
\label{fig:dense} 
\end{figure} 
The overall architecture of our model is illustrated in Fig.~\ref{fig:overview}. We concatenate patches from FLAIR and T2, and concatenate patches from native (T1) and post contrast T1-weighted (T1-CE) to generate separate inputs for two independent feature extractors, which share the same structure. The top pathway in Fig.~\ref{fig:overview} performs a binary classification that segments the whole tumor from the background. The bottom pathway in Fig.~\ref{fig:overview} concatenates features extracted from T1 and T1-CE with those from FLAIR and T2 (extracted by the top pathway), then feeds them into a 4-class softmax classifier (e.g., background, ED, ET and NCR/NET). Both pathways are updated during the training phase together. In the testing phase, the final predictions are generated from the bottom pathway. This hierarchical pipeline boosts the performance since it follows the label hierarchy of lesion regions. The FLAIR and T2 scans are mainly responsible for separating out ED regions, which are always the largest lesion areas. The ET and NCR/NET regions, segmented with T1 and T1-CE, are inside of ED regions. We take $38 \times 38 \times 38$ patches as inputs and predict the center $12 \times 12 \times 12$ voxels simultaneously at each iteration.

\subsection{Two-Stage Densely Connected 3D CNN}

Detailed information of our feature extractors is provided in Table~\ref{table:feature_extractor}. We use 24 kernels with each kernel having size of $3^3$ for the initial convolution. In stage 1 and 2, two densely connected blocks are utilized respectively. An illustration is shown in Fig.~\ref{fig:dense}. Each block has 6 convolutional layers with the growth rate of 12. A densely connected block is formulated as: 
\begin{align}
x_{l + 1} & = \mathcal{H}([x_{0}, x_{1}, ..., x_{l}]) \enspace,\\
\mathcal{H}(x_i) & = W*\sigma(B(x_i)) \enspace, 
\enspace i \in \{0, ..., l\} \enspace,
\end{align}
where $W$ is the weight matrix, $*$ denotes convolution operation, $B(\cdot)$ represents batch normalization~\cite{DBLP:conf/icml/IoffeS15}, $\sigma(\cdot)$ denotes rectified linear unit~\cite{ReLU} for activation and $[x_{0}, x_{1}, ..., x_{l}]$ represents the concatenation of all outputs of previous layers before the layer of $l+1$. The feature dimension $d_{l}$ of $x_{l}$ is calculated as: 
\begin{equation}
d_{l} = d_{0} + g \cdot l \enspace,
\end{equation}
where $g$, the \textit{growth rate}, is the number of kernels used in each convolutional layer, and $d_0$ is the feature dimension after the initial convolution.
\begin{table}
\centering
\begin{tabular}{ |c|c|c|c| }
\hline
Stage & Layers & \# Features. & Recep. Field \\
\hline
Convolution & $3^3 \times 24$ conv. & 24 & $3^3$\\
\hline
Stage 1 & [$3^3 \times 12$ conv.] $\times$ 6 & 96 & $15^3$ \\
\hline
Stage 2 & [$3^3 \times 12$ conv.] $\times$ 6 & 168 & $27^3$ \\
\hline
\end{tabular}
\caption{Specifications of the two-stage feature extractor. A densely connected block is applied at each stage. }
\label{table:feature_extractor}
\end{table}
Receptive field is a region of the input data that is involved in the calculation of a node in hidden layers of a CNN. 
The size of receptive field of a particular node, which is the output of a convolutional layer without pooling, can be obtained by: 
\begin{equation}\label{eq:rf_size}
    r_{l+1} = r_{l} + (k - 1) \cdot s_{l} \enspace,
\end{equation}
where $l$ is the layer index, $k$ is the kernel size, and $s_{l}$ is the current stride size (we set as 1).   
In our model, we do not use any downsampling layers in order to preserve the high resolution of patches. Thus, the receptive fields of outputs of the two stages are 15 and 27, respectively, in each dimension of a 3D kernel.

An $1^3 \times 168$ convolution operation follows the dense block of each stage. Due to the concatenation operations in dense blocks, features from the lowest to the highest level persist. The post $1^3$ convolution serves the purpose of merging features from all levels of abstraction. We have the number of kernels the same as that of the output of the dense block.


\section{EXPERIMENT AND RESULTS}
\subsection{Data}

One of the challenges in working with MRI data is the dealing with artifacts produced by inhomogeneity in the magnetic field, which is called bias field distortion. This type of distortion is prevalent in MRI scans, which may jeopardize the performance of CNN models. We employ N4ITK bias correction~\cite{DBLP:journals/tmi/TustisonACZEYG10} for all MRI sequences. This algorithm removes the intensity gradient of each scan. After normalizing the MRI sequences with N4ITK, we normalize each MRI sequence by subtracting the mean value and dividing by its standard deviation. 

We use the training data of BraTS 2017 challenge, which includes MRI scans of 210 patients, to train and test our model. This dataset is challenging since we are interested in segmenting not only the complete tumor, but also subregions of the tumor, e.g., the necrotic and non-enhancing tumor (NCR/NET), peritumoral edema (ED) and GD-enhancing tumor (ET). Fluid Attenuated Inversion Recovery (FLAIR), native (T1), post contrast T1-weighted (T1-CE) and T2-weighted (T2) scans are acquired for every subject. Each of these modalities captures different properties of the complete tumor and, hence, provides additional useful information required to segment gliomas subregions. We randomly split the whole dataset into training and testing sets, each with 168~(80\%) and 42~(20\%) MRI sequences. We do not have separate validation set due to the limited amount of data. Instead, we apply the 4-fold validation procedure to tune hyper-parameters. We generate the training patches by the following two steps: (1) randomly sample voxels with 50\% being in lesion regions; (2) extract $38 \times 38 \times 38$ 3D patches and concatenate these patches from FLAIR and T2, and from T1 and T1-CE to obtain two $38 \times 38 \times 38 \times 2$ tensors as the inputs to our two-pathway architecture.

\subsection{Training and Testing}
We use the cross entropy as the loss function, which is formulated as: 
\begin{equation}
\text{CE}(y^{(i)}, z^{(i)}) = -\sum_{j=1}^{k}1\{y^{(i)} = j\} \text{log}\frac{e^{z_j}}{ \sum_{l=1}^{k} e^{z_l}}
\enspace,
\end{equation}
where $y^{(i)}$ is the ground-truth of the $i^{th}$ training sample, $z^{(i)}$ is the score output by the model and $k$ is the number of classes. 

For each training epoch, we randomly extract 400 $38^3$ patches with 200 patches centered on the lesion area, and 200 patches centered on the background. Adam optimizer~\cite{DBLP:journals/corr/KingmaB14} is applied with the initial learning rate of 5e-4. Our model is implemented in Tensorflow~\cite{tensorflow}. It takes 2.5 hours to train for an epoch on a single NVIDIA GTX 1080 GPU. The network converges after approximately 5 epochs. 

To test our model, we first pre-process each testing MRI sequence with N4ITK and normalization. Then, we feed into the network the non-overlapping $38^3$ patches extracted from each MRI sequence to obtain the prediction of $12^3$ at the center of patches. We use sliding window approach to obtain a dense prediction for an MRI sequence.

\subsection{Results}

\subsubsection{Metric}
We report the testing results in terms of \textit{Dice}, a widely used metric for evaluating lesion segmentation models. Dice score is calculated as: 
\begin{equation}
\text{Dice} = \frac{2\text{TP}}{\text{FN}+\text{FP}+2\text{TP}}
\enspace,
\end{equation}
where TP, FP and FN stands for ``true positive'', ``false positive'' and ``false negative'' predictions, respectively. It measures the overlap area of the predicted lesion region and the ground-truth. 

\subsubsection{Qualitative Results}


\begin{figure}[!b]
\centering
\includegraphics[width=0.9\linewidth]{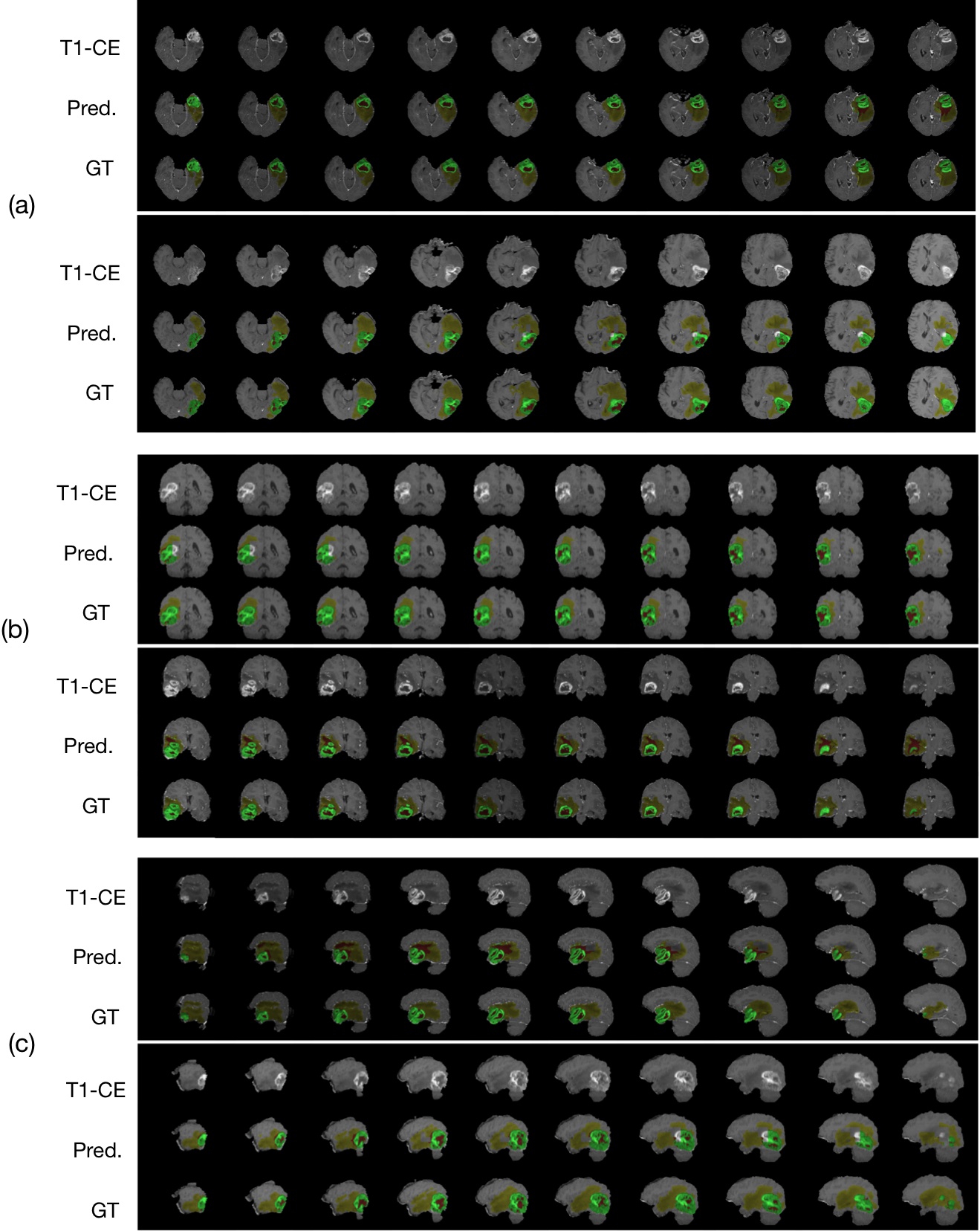}
\caption{Randomly selected segmentation results of two MRI sequences. T1-CE scans, segmentations and the ground-truth are presented at the top, middle and bottom, respectively. The complete tumor, tumor core and enhancing tumor are indicated in yellow, green and red masks, respectively. (a), (b) and (c) are the slices from the axial plane, coronal plane and sagittal plane, respectively.}
\label{fig:visualize_1}
\end{figure}


We randomly select two MRI sequences (Brats17\_TCIA\_121\_1 and Brats17\_TCIA\_377\_1) and visualize the segmentation results in Fig.~\ref{fig:visualize_1}. The 2D MRI scan slices on the axial plane view, coronal plane and sagittal plane view are visualized in Fig.~\ref{fig:visualize_1}(a), Fig.~\ref{fig:visualize_1}(b) and Fig.~\ref{fig:visualize_1}(c), respectively. From these example segmentations, we notice that our model has a promising performance for 2D MRI slices. Figure~\ref{fig:visualize_1}(a) suggests that our model has the ability to accurately separate out lesion subregions. The segmentation results of the tumor core and enhancing tumor are most impressive. One limitation of our model, from our observation, is that the predictions are inconsistent along the axis perpendicular to the axial plane. In most cases, lesion regions are concentrated in the center of the whole MRI sequence. However, our model predicts that benign regions can be surrounded by lesion regions in Fig.~\ref{fig:visualize_1}(b) and Fig.~\ref{fig:visualize_1}(c) , which is unlikely to happen. One possible reason is that the 3D convolution cannot capture the long-term dependencies in MRI sequences due to our patch-wise training schema. Our future goal is to address this inconsistency in our predictions.

\subsubsection{Quantitative Results}

Table~\ref{table:results} shows the performance of three state-of-the-art models and ours. The prediction of four subregions is aggregated to generate the ``complete tumor''~(label 1,2,4), ``tumor core''~(label 1,4) and ``enhancing tumor''~(label 4). Both Pereira et al.~\cite{DBLP:journals/tmi/PereiraPAS16} and Zhao et al.~\cite{DBLP:conf/miccai/ZhaoWSLFZ16} apply 2D CNN models that take 2D $33 \times\ 33$ patches as inputs and predict the label of the center voxel. These architectures can be substantially inefficient due to the repeated computation. Zhao et al.~\cite{DBLP:conf/miccai/ZhaoWSLFZ16} further apply conditional random field~(CRF) to refine the predictions given by the CNN. The model proposed by Kamnitsas et al.~\cite{DBLP:journals/mia/KamnitsasLNSKMR17} makes dense predictions according to the receptive field of each voxel in the final feature map, which inspired the architecture of this work. They build two almost identical pathways that take the original MRI scans and downsampled ones into account, claiming that the use of low-resolution inputs helps incorporate larger contextual information. In our work, instead of modifying the inputs, we incorporate multi-scale contextual information by making predictions at different depths of the network, which makes our model incredibly efficient and boosts the performance.
\begin{table}[ht!]
\centering
\begin{tabular}{c|c|c|c}
\toprule
\hline
\rule[-1ex]{0pt}{3.5ex} & Comp. & Core & Enh. \\
\hline 
\rule[-1ex]{0pt}{3.5ex} Pereira et al.~\cite{DBLP:journals/tmi/PereiraPAS16} & 0.84 & 0.72 & 0.62 \\
\rule[-1ex]{0pt}{3.5ex} Kamnitsas et al.~\cite{DBLP:journals/mia/KamnitsasLNSKMR17} & \textbf{0.90} & 0.75 & 0.73 \\
\rule[-1ex]{0pt}{3.5ex} Zhao et al.~\cite{DBLP:conf/miccai/ZhaoWSLFZ16} & 0.87 & \textbf{0.83} & 0.76 \\
\rule[-1ex]{0pt}{3.5ex} Ours & 0.72  & \textbf{0.83} & \textbf{0.81} \\
\bottomrule
\end{tabular}
\caption{The Dice scores of selected existing models and the proposed model are presented in this table. Our model has the highest dice score in tumor core and enhancing tumor segmentation.}
\label{table:results}
\end{table} 
The proposed architecture achieves Dice scores of 0.72, 0.83 and 0.81 for the complete tumor, tumor core and enhancing tumor segmentation, respectively. Compared with previous models, our method achieves the highest Dice score in tumor core and enhancing tumor segmentation. Kamnitsas et al.~\cite{DBLP:journals/mia/KamnitsasLNSKMR17} obtain the highest score in the complete tumor segmentation. These results demonstrate the potential of our model in 3D MRI segmentation tasks.

\subsubsection{Ablation Study}
\begin{table}[ht]
\centering
\begin{tabular}{ c|c|c|c }
\toprule
\hline
\rule[-1ex]{0pt}{3.5ex} & Comp. & Core & Enh.  \\
\hline 
\rule[-1ex]{0pt}{3.5ex} Non-Dense (See Tab.~\ref{table:ab_nondense}) & 0.61  & 0.77 & 0.78 \\
\rule[-1ex]{0pt}{3.5ex} Non-Hierarchical & \textbf{0.74}  &  0.74 & 0.75 \\
\rule[-1ex]{0pt}{3.5ex} Single-Scale Recep. Field & 0.62  & 0.64 & 0.62   \\
\rule[-1ex]{0pt}{3.5ex} Proposed & 0.72 & \textbf{0.83} & \textbf{0.81}\\ 
\bottomrule

\end{tabular}
\caption{The experiment results of our ablation study. We separately substantiate the effectiveness of densely connected block, hierarchical inference structure, and multi-scale receptive fields. It can be justified that the removal of each component impairs the overall performance to some extent. } 
\label{table:result_ablation}
\end{table} 




\noindent \textbf{Non-Dense Structure.} \quad To indicate the improved performance of the dense structure, we build 6 convolutional layers (see Tab.~\ref{table:ab_nondense}) in the linear chain topology for each stage so that the size of the receptive field corresponds to that of the proposed model. Each upper convolution layer has 12 more kernels than the lower layer, which makes the number of features at the predictions layers equal to that of its counterpart. Other components of this non-dense architecture are the same with the proposed model.

\noindent \textbf{Non-Hierarchical Structure.} \quad To understand the effectiveness of the hierarchical inference structure, we discard the two pathways pipeline illustrated in Fig~\ref{fig:overview}. We concatenate all four types of scans as the inputs, i.e., we extract $38 \times 38 \times 38 \times 4$ patches to feed into the network. The classification layers directly predict the scores for four labels. Other configurations remain the same as the proposed model. 



\begin{wraptable}{l}{0.49\linewidth}
\begin{center}
\begin{tabular}{ c | c }
\toprule
\hline
Stage &Layers \\
\hline
Convolution & $3^3 \times 24$ conv. \\
\hline
\multirow{6}{*}{Stage 1}&$3^3 \times 36$ conv. \\
\hhline{~-}
& $3^3 \times 48$ conv.\\
\hhline{~-}
&$3^3 \times 60$ conv.\\
\hhline{~-}
&$3^3 \times 72$ conv.\\
\hhline{~-}
&$3^3 \times 84$ conv.\\
\hhline{~-}
&$3^3 \times 96$ conv.\\
\hhline{--}

\multirow{6}{*}{Stage 2}& $3^3 \times 108$ conv.\\
\hhline{~-}

&$3^3 \times 120$ conv.\\
\hhline{~-}
&$3^3 \times 132$ conv.\\
\hhline{~-}
&$3^3 \times 144$ conv.\\
\hhline{~-}
&$3^3 \times 156$ conv.\\
\hhline{~-}
&$3^3 \times 168$ conv.\\
\hhline{--}
\bottomrule
\end{tabular}
\end{center}
\caption{Specifications of the two-stage feature extractor with traditional convolutional layers (Non-Dense). This feature extractor has the same number of features as the densely connected counterpart at prediction layers. }
\label{table:ab_nondense}
\vspace{-3mm}
\end{wraptable}

\noindent \textbf{Single-Scale Recep. Field Structure.} \quad To verify the advantage of using multi-scale receptive fields, we remove all layers in stage 2 and make predictions only with the output of stage 1. We still make the model predict $12^3$ in one iteration, so the dimension of the input patches becomes $26 \times 26 \times 26 \times 2$. Densely connected blocks and the hierarchical inference structure are still applied. 


These ablation study results (see Tab.~\ref{table:result_ablation}) substantiate the effectiveness of each component of the proposed model. The non-hierarchical model achieves the closest overall performance to the proposed one. However, it is still outperformed by the proposed model by a large margin in tumor core and enhancing tumor segmentation. The non-dense counterpart is outperformed by the proposed model in all three categories. It suggests the effectiveness of densely connected blocks is obvious even though the network is relatively shallow compared with those ultra deep networks used in natural image recognition. It is worthwhile to mention that the non-dense network is rather inefficient as more parameters are introduced. The model with single-scale receptive field has the lowest accuracy. It further demonstrates that taking into account different levels of contextual information significantly boost the classification accuracy. \\


\section{ DISCUSSION AND CONCLUSION}

In this paper, we introduce a new approach for brain tumor segmentation in MRI scans. DenseNet was initially introduced for the image classification problem. In this work, we explore the potential of densely connected blocks in 3D segmentation tasks. Compared with traditional networks with no skip connections, the improved information flow extracts better features and significantly help the optimization. We take multi-scale receptive fields into account to accurately classify voxels. Our model is made to predict $12^3$ voxels in one iteration to increase the efficiency. Experimental results indicate that the proposed model performs on par with the state-of-the-art models without advanced tricks. The ablation study substantiates the effectiveness of each component of the whole architecture. We thus conclude that the proposed model has great potential for MRI segmentation or other medical image segmentation tasks.

{\small
\bibliography{report} 
\bibliographystyle{spiebib} 
}

\end{document}